\documentclass[runningheads]{llncs}
\usepackage[T1]{fontenc}
\usepackage{graphicx}
\usepackage{amsmath,amssymb,amsfonts}
\usepackage{subcaption}
\usepackage{booktabs}
\usepackage{multirow}

\begin{document}
\title{SynthVC: Leveraging Synthetic Data for End-to-End Low Latency Streaming Voice Conversion}
%
%
\author{Zhao Guo \and
Ziqian Ning \and
Guobin Ma \and
Lei Xie
}
\authorrunning{Zhao Guo et al.}
%
\institute{Audio, Speech and Language Processing Group (ASLP@NPU), School of Software, Northwestern Polytechnical University, Xi'an, China \\
\url{http://www.npu-aslp.org/}\\
\email{gzhao@mail.nwpu.edu.cn}}
\maketitle              
\begin{abstract}
Voice Conversion (VC) aims to modify a speaker’s timbre while preserving linguistic content. While recent VC models achieve strong performance, most struggle in real-time streaming scenarios due to high latency, dependence on ASR modules, or complex speaker disentanglement, which often results in timbre leakage or degraded naturalness. We present SynthVC, a streaming end-to-end VC framework that directly learns speaker timbre transformation from synthetic parallel data generated by a pre-trained zero-shot VC model. This design eliminates the need for explicit content–speaker separation or recognition modules. Built upon a neural audio codec architecture, SynthVC supports low-latency streaming inference with high output fidelity. Experimental results show that SynthVC outperforms baseline streaming VC systems in both naturalness and speaker similarity, achieving an end-to-end latency of just 77.1 ms. 

\end{abstract}

\section{Introduction}
Voice conversion (VC), the technique of converting speaker timbre while preserving linguistic content~\cite{sisman2020overview}, has achieved significant progress through deep learning advancements. Modern VC systems demonstrate remarkable capabilities to achieve both speaker similarity and speech naturalness, enabling applications ranging from movie dubbing~\cite{ning2023expressive,yao2023preserving} to voice privacy protection~\cite{yao2023distinguishable}. Conventional VC approaches~\cite{li2023freevc,hussain2023ace,wang2021enriching,li2021starganv2} typically operate on complete utterances, requiring full-sentence input to generate converted speech. While effective for offline conversion, this utterance-level paradigm faces critical limitations in real-time communication (RTC) scenarios such as live streaming and video conferencing, where streaming processing with strict latency constraints is essential.

Streaming voice conversion introduces unique technical challenges due to its causal processing requirements. Unlike non-streaming models, the causal processing constraint requires frame-level or chunk-wise input handling with strictly limited access to future context. The absence of future context results in degraded performance, including relatively lower intelligibility, poorer sound quality, and inferior speaker similarity. On the other hand, the causal model design and caching to ensure output continuity during streaming inference introduce additional complexity to streaming voice conversion.

With limited future information in streaming voice conversion, the shortcomings of existing disentangling approaches are magnified. The mainstream approach to disentanglement is to use an automatic speech recognition (ASR) model to extract speaker-independent bottleneck features (BNF) as input to the VC model~\cite{ning23_interspeech,Ning2023Dualvc2D,streamvoice}. While this approach benefits from semantic-rich BNF features, three fundamental limitations exist under the streaming model setup: (1) Performance degradation of streaming ASR models leads to potential timbre leakage in BNF which causes trade-offs between naturalness and speaker similarity; (2) The inherent latency requirements of streaming ASR models (typically requiring tens to hundreds of milliseconds lookahead), fundamentally constrain minimum achievable system delay; (3) Cascaded processing introduces error propagation and complex system pipeline. 
As an alternative to ASR-based feature extraction, speech representation disentanglement (SRD) methods aim to separate content and speaker information through model structure design or tailored training losses, without relying on external feature extractors.
These approaches typically employ methods such as mutual information minimization~\cite{wang2021vqmivc}, gradient reversal~\cite{wang2021adversarially}, or information bottlenecks~\cite{qian2019autovc,ning24_interspeech,ma2024vec,zhu2025zsvc} to disentangle speaker information from the linguistic content. However, such disentanglement methods often struggle under streaming constraints, as they require carefully tuned model structures to maintain the trade-off between speaker similarity and naturalness. 
Instead of continuing to adapt disentanglement strategies to voice conversion, we pursue an alternative direction: bypassing disentanglement entirely by enabling supervised training through synthetic parallel data constructed from non-parallel corpora.

While several streaming VC systems~\cite{Ning2023Dualvc2D,ning23_interspeech,chen2023streaming,hayashi2022investigation,an2022cuside} employ knowledge distillation to mitigate quality degradation caused by streaming constraints, these methods often inherit the limitations of upstream models and introduce considerable system complexity.
Rather than further adapting disentanglement-based methods for streaming scenarios, we pursue a different direction: adopting a neural codec architecture originally developed for low-latency audio compression~\cite{zeghidour2021soundstream,defossez2022encodec,wu2023audiodec}, which naturally supports streaming processing while enabling high-quality speech generation.

In this work, we present SynthVC, a streaming end-to-end voice conversion framework that performs direct speaker timbre mapping in the latent space of an autoencoder. Built on AudioDec~\cite{wu2023audiodec}, SynthVC supports efficient waveform-to-waveform conversion with native streaming capability. To enable supervised training without relying on ASR models or disentanglement mechanisms, we adopt a pre-trained zero-shot VC model (Seed-VC~\cite{liu2024seedvc}) as a synthetic parallel data generator, allowing supervised training with diverse timbre mappings. Our audio samples are available~\url{https://anonymous.4open.science/w/SynthVC-BD0D/}.


\section{Related Work}

Voice conversion has seen rapid progress across multiple research directions. This section reviews two lines of work that are most relevant to our method: zero-shot voice conversion models, which eliminate the need for speaker-content disentanglement, and neural audio codecs, which provide low-latency and high-quality waveform modeling. 

\subsection{Zero-Shot Voice Conversion}

Zero-shot voice conversion (VC) aims to convert speech to match the
timbre of any unseen speaker, given only a short reference utterance. This setting is particularly attractive for  flexible and
generalizable VC systems, where new speakers can be supported at inference time without fine-tuning.

Early approaches such as AutoVC~\cite{qian2019autovc} and YourTTS~\cite{DBLP:conf/icml/CasanovaWSJGP22} rely on speaker-independent content encoders and global speaker embeddings. While effective, these methods often struggle with timbre leakage, where residual source timbre contaminates the converted speech, and with degraded intelligibility due to overly aggressive content bottlenecks.

More recent methods have explored diffusion-based generation and large-scale training to improve generalization and audio quality. Seed-VC~\cite{liu2024seedvc} addresses both timbre leakage and training-inference mismatch by introducing a timbre shifter during training and employing a diffusion transformer with in-context learning. This allows Seed-VC to capture fine-grained speaker attributes from full reference utterances and achieve state-of-the-art performance in both speaker similarity and word error rate (WER).

Unlike prior works that optimize zero-shot VC performance, we repurpose Seed-VC as a parallel data generator for supervised training.

\subsection{Neural Audio Codecs} 

Neural audio codecs have recently emerged as an effective foundation for real-time speech generation tasks due to their ability to perform high-quality waveform reconstruction under low-latency, streamable conditions. Compared with traditional vocoders or parametric codecs, neural codecs such as SoundStream~\cite{zeghidour2021soundstream}, EnCodec~\cite{defossez2022encodec}, and AudioDec~\cite{wu2023audiodec} offer greater fidelity and runtime efficiency, making them attractive backbones for streaming voice conversion (VC).

Recent work such as StreamVC~\cite{yang2024streamvc} demonstrates that codec-based architectures can enable real-time VC with end-to-end latency as low as 70.8 ms on mobile devices. StreamVC leverages SoundStream as its decoder backbone and incorporates HuBERT-derived soft units, whitened fundamental frequency (F0), and a causal convolutional decoder to achieve high pitch stability and intelligibility. However, the reliance on externally extracted features introduces additional latency and potential speaker leakage, and the system complexity remains high due to multi-stream conditioning.

In contrast, our work builds on AudioDec~\cite{wu2023audiodec}, a modular neural codec designed for real-time speech synthesis. AudioDec features a causal encoder–quantizer–decoder architecture, integrates a HiFi-GAN-based vocoder with multi-period discriminators, and supports sub-10 ms inference latency even on CPUs.
Leveraging its streamable structure, we develop a lightweight VC model capable of low-latency waveform-level conversion without relying on external linguistic features.

\section{Methodology}

\label{sec:method}

\subsection{Overview}

We propose SynthVC, a framework that combines the low-latency, high-fidelity properties of neural codecs with end-to-end training using synthetic parallel data.

SynthVC builds upon AudioDec~\footnote{https://github.com/facebookresearch/AudioDec}, an open-source neural codec architecture designed for streamable speech generation. We extend its modular framework by inserting a speaker transformation module between the encoder and decoder, enabling speaker-conditioned latent-to-latent conversion. The architecture retains three fundamental components: an \textit{autoencoder} for latent space modeling, a \textit{converter} for speaker transformation, and \textit{discriminators} for quality enhancement.

To supervise the training of SynthVC without requiring parallel data, we utilize a high-quality zero-shot VC model to construct synthetic parallel waveform pairs. These data simulate conversions across diverse speakers and allow the converter to learn precise timbre mappings in a supervised manner. The full pipeline and training strategy are detailed in the following sections.

\subsection{Parallel Dataset Construction}

We adopt Seed-VC\footnote{https://github.com/Plachtaa/seed-vc}, a recent zero-shot VC model, as our synthetic data generator. Seed-VC accepts a source waveform and a reference waveform, and outputs a converted version that transfers the reference speaker's timbre while preserving the source content. This allows us to construct high-quality parallel pairs from non-parallel corpora.

Given an original waveform $w$ and a randomly selected reference waveform $w_{\text{ref}}$ from a timbre-diverse pool, we use Seed-VC to generate the converted waveform:

\begin{equation}
w_{\text{syn}} = T(w, w_{\text{ref}})
\end{equation}

We retain the speaker ID $sid$ associated with $w$ as the target label, resulting in training triplets $(w_{\text{syn}}, w, sid)$. This setup enables the model to learn to reverse the timbre shift introduced by the generator and recover the original speaker characteristics. By sampling diverse references for each source utterance, we simulate many-to-one conversions and enrich training diversity.

\begin{figure*}[h]
  \centering
  \includegraphics[width=\linewidth]{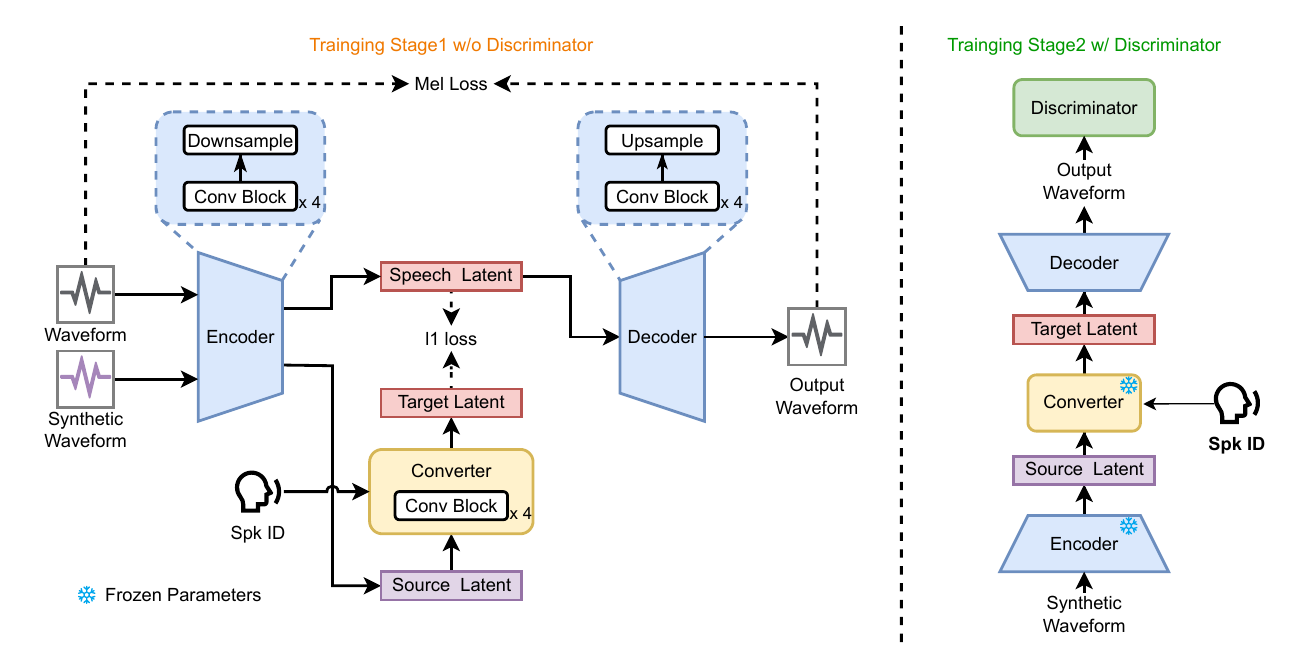}
  \caption{The overall framework of SynthVC consists of a two-stage training strategy.} 
  \label{fig:DistilVC}
\end{figure*}

\subsection{Model Architecture}

As illustrated in Figure~\ref{fig:DistilVC}, SynthVC builds upon the modular autoencoder architecture of AudioDec, 
by omitting AudioDec’s original vector quantization layer to support continuous latent representations and fully differentiable training.
SynthVC comprises three main components: an autoencoder for speech representation learning, a timbre converter for speaker transformation, and a discriminator module for adversarial enhancement.

\begin{figure}
    \centering
    \includegraphics[width=1\linewidth]{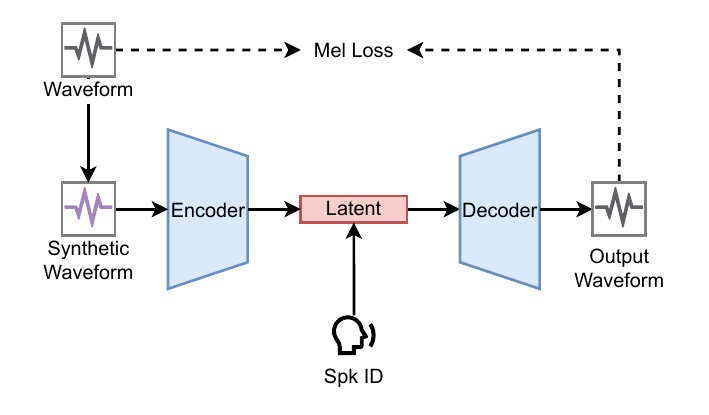}
    \caption{Waveform-level training uses synthetic waveforms as input and original waveforms as supervision. This approach can lead to over-smoothing and loss of audio detail.}
    \label{fig:waveform-supervision}
\end{figure}

As illustrated in Figure~\ref{fig:waveform-supervision}, a straightforward strategy for voice conversion is to train the autoencoder using parallel waveform pairs, where the input is the source speaker’s utterance and the target is a converted utterance with the same linguistic content but a different speaker timbre. However, our experiments show that this approach leads to over-smoothing, where the reconstructed audio lacks high-frequency detail and sounds muffled or unnatural. We attribute this to the fact that mel-spectrogram-based L1 losses encourage the model to average across variations in speaker timbre, especially when multiple speaker identities are involved. This results in blurry reconstructions and makes it difficult for the model to explicitly represent speaker-specific information in the latent space.

To overcome this limitation, we propose a latent-to-latent conversion framework. In our design, the autoencoder first learns to encode speech into a latent representation that captures both content and acoustic detail. A dedicated Converter module is then introduced between the encoder and decoder. This module transforms the source latent into a target latent conditioned on the target speaker identity. By functionally separating compression and timbre transformation, we allow the encoder to focus on capturing content and acoustic details, while the converter learns speaker-specific mappings. This design improves reconstruction fidelity.

\noindent\textbf{Autoencoder}\quad
The autoencoder follows the encoder–decoder design of AudioDec. The encoder comprises four convolutional blocks, each with three residual units and a downsampling layer. The decoder mirrors this structure with corresponding upsampling modules. It maps latent representations back to audio waveforms.

\noindent\textbf{Converter}\quad
The Converter consists of four convolutional blocks, each with three residual units. It takes the source latent and a learnable embedding of the target speaker ID as input, and outputs a transformed latent that matches the target speaker’s timbre. By operating entirely in the latent space, the converter avoids entanglement with waveform-level distortions and enables finer control over speaker characteristics. Our experiments show that this design improves high-frequency spectral detail compared to waveform-level conversion alone.

\noindent\textbf{Discriminator}\quad
To further improve naturalness, we adopt the adversarial architecture from UnivNet~\cite{jang2021univnet}, incorporating multi-resolution spectrogram discriminators (MRSD) and multi-period discriminators (MPD). These discriminators guide the decoder to produce high-fidelity speech with better periodic structure and spectral consistency.

\subsection{Training Strategy}

During training stage 1, we train both the Autoencoder and Converter using only the metric loss, which ensures rapid and stable convergence. The Encoder encodes \( w \) into the latent representation \( z \), which is then reconstructed into the waveform \( \hat{w} \) by the Decoder. For a synthetic waveform \( w_{syn} \), we reuse the Encoder to extract its source speech latent representation \( z_{src} \). To ensure the converted audio retains the high-frequency acoustic details, we convert the latent representation rather than the waveform itself. The Converter then utilizes the speaker ID as a global condition to learn the mapping from the source speech latent representation \( z_{src} \) to the target speaker latent representation \( z_{tgt} \). 

The mel loss measures the distance between the mel spectrograms of the output waveform \(\hat{w}\) and the real waveform \(w\), calculated as:

\begin{equation}
L_{\text{mel}} = \mathbb{E} \left[ \lVert\text{mel}(w) - \text{mel}(\hat{w})\rVert_1 \right]
\end{equation}
where \( mel() \) denotes the mel spectrogram extraction operation. The conversion loss measures the distance between the speech latent representation \( z \) and the target latent representation \( z_{tgt} \), and is  computed using L1 loss:

\begin{equation}
L_{\text{conv}} = \mathbb{E} \left[ \lVert{z} - z_{tgt}\rVert_1 \right]
\end{equation}
The training objectives at this stage include the mel loss \( L_{mel} \) and conversion loss \( L_{conv} \), with the total loss function defined as:

\begin{equation}
L = a \cdot L_{mel} + b \cdot L_{conv}
\end{equation}
where \( a \) and \( b \) are the weights for the mel loss and conversion loss, set to 45 and 5, respectively. 

Through the joint training of the Autoencoder and Converter, the first stage enables end-to-end conversion from the source speech to the target speaker.

During training stage 2, we introduce a generative adversarial network~\cite{goodfellow2020generative} (GAN). The Discriminators are jointly trained with only the Decoder to enhance reconstruction quality, focusing on waveform details and phase synchronization.

To ensure the Decoder receives consistent latent representations during training and inference, we adopt the \textit{Aligned Training} strategy. Specifically, we use the Converter's output as the input to the Decoder during training, instead of the Encoder's output. This prevents distribution mismatch and helps the model generate clearer and more stable speech.

We extract the latent representation \( z \) through the Encoder and Converter. The Decoder, denoted as \( G \), reconstructs \( z \) back into the waveform. The mel loss in stage 2 is defined as:

\begin{equation}
L_{mel}' = \mathbb{E} \left[ \lVert \text{mel}(w) - \text{mel}(G(z)) \rVert_1 \right]
\end{equation}
The discriminator is denoted as \( D \). The adversarial loss for the generator \( G \) and the discriminator \( D \) is given by:

\begin{align}
L_{adv}(D) &= \mathbb{E}_{(w,z)} \left[ (D(w) - 1)^2 + (D(G(z)))^2 \right] \\
L_{adv}(G) &= \mathbb{E}_{z} \left[ (D(G(z)) - 1)^2 \right]
\end{align}
Additionally, the feature matching loss is expressed as:

\begin{equation}
L_{fm}(G) = \mathbb{E}_{(w,z)} \left[ \sum_{l=1}^{T} \frac{1}{N_l} \lVert D^l(w) - D^l(G(z)) \rVert_1 \right]
\end{equation}

The total loss for the generator in the second stage is given by:

\begin{equation}
L(G) = a \cdot L_{mel}' + c \cdot L_{adv}(G) + d \cdot L_{fm}(G)
\end{equation}
where \( a \), \( c \), and \( d \) represent the weights for the mel loss, adversarial loss, and feature matching loss, set to 45, 1, and 2, respectively.

\section{Experimental Setup}

\subsection{Dataset}

We use the open-source Mandarin corpus Aishell3~\cite{shi2020aishell} as our main dataset, which contains 88,035 samples from 218 speakers. We reserve 100 samples for testing. All audio is resampled to 16 kHz. During evaluation, 10 target speakers are randomly selected from Aishell3, and all test utterances are converted to these speakers. Additionally, we randomly sample 400,000 utterances from the Emilia dataset~\cite{emilia,amphion} as the reference corpus to enhance speaker diversity during synthetic data generation.

\subsection{Synthetic Data Generation}

To generate training data, we use Seed-VC~\cite{liu2024seedvc}, a high-quality zero-shot voice conversion model. We follow its recommended inference settings: \texttt{inference-cfg-rate}=0.7, \texttt{auto-f0-adjust}=True, and \texttt{length-adjust}=1.0. The number of diffusion steps is randomly sampled between 10 and 25 to balance quality and throughput. For each utterance in the Aishell3 training set, six reference utterances are randomly sampled from the reference corpus to produce six synthetic converted versions, forming synthetic parallel training pairs.

\subsection{Training Configuration}

All models are trained on a single NVIDIA RTX 4090D GPU with a batch size of 16. Each utterance is segmented into 1-second chunks. The training consists of two stages: in the first 200k steps, we jointly optimize the encoder, decoder, and converter using reconstruction and latent alignment losses. In the second stage (200k–700k steps), the encoder and converter are frozen, and the decoder is fine-tuned with adversarial losses using the converted latent representation to ensure training–inference consistency.

\subsection{Baselines and Variants}
We compare our proposed SynthVC with the following baseline models:

\begin{itemize}
  \item \textbf{DualVC2}: an ASR-based VC model using bottleneck features for disentanglement.
  \item \textbf{DualVC3}: a streaming VC model trained with SRD-based techniques. We run it in stand-alone mode (without the language model) to reduce latency.
  \item \textbf{Seed-VC}: a diffusion-based zero-shot VC model, used as a generator in our system. Although non-streamable, its performance represents a quality upper bound not constrained by real-time requirements.
\end{itemize}

We also compare three configurations of SynthVC: small, base, and large, each using different latent dimensions and network widths. Their computational costs are listed in Table~\ref{config}.

\subsection{Evaluation Metrics}
We evaluate all models using both subjective and objective metrics:

\begin{itemize}
    \item Subjective Evaluation: We conduct 5-point MOS tests on naturalness (N-MOS), speaker similarity (S-MOS), and intelligibility (I-MOS), using 20 native Mandarin speakers per sample.
    \item Objective Evaluation: We use the \texttt{seed-tts-eval} toolkit\footnote{\url{https://github.com/BytedanceSpeech/seed-tts-eval}} for two objective metrics. To evaluate intelligibility, we compute the Character Error Rate (CER) using Paraformer-zh, a Mandarin ASR model. For speaker similarity, we calculate the Speaker Cosine Similarity (SPK-COS) by extracting speaker embeddings with a WavLM-large model fine-tuned for speaker verification and measuring the cosine similarity between converted and reference utterances.
\end{itemize}

\begin{table*}[ht]
  \caption{Subjective evaluation results in terms of 5-point MOS for naturalness (N-MOS), speaker similarity (S-MOS), and intelligibility (I-MOS), with 95\% confidence intervals.}
  \label{tab:subjective}
  \centering
  \resizebox{0.7\textwidth}{!}{
    \begin{tabular}{ l c c c }
      \toprule
      \textbf{Model} & \textbf{N-MOS$\uparrow$} & \textbf{S-MOS$\uparrow$} & \textbf{I-MOS$\uparrow$} \\
      \midrule
      ground-truth              &  4.43±0.04    & N/A       & 4.56±0.03    \\
      Seed-VC                   & \textbf{4.02±0.06}    & \textbf{4.34±0.05} & \textbf{4.41±0.05}    \\
      DualVC2                   &  3.41±0.05    & 3.65±0.06 & 3.78±0.04    \\
      DualVC3(stand-alone mode) &  3.19±0.08    & 3.57±0.04 & 3.46±0.06    \\
      \midrule
      SynthVC-large             & \textbf{3.68±0.06}     & \textbf{3.95±0.06} & \textbf{3.85±0.06}    \\
      SynthVC-small             & 3.46±0.10     & 3.72±0.12 & 3.53±0.05    \\
      SynthVC-base              & 3.51±0.09     & 3.77±0.08 & 3.76±0.05    \\
      \bottomrule
    \end{tabular}}
\end{table*}

\begin{table*}[ht]
  \caption{Objective evaluation results including character error rate (CER), speaker cosine similarity (SPK-COS), and total streaming latency (Latency).}
  \label{tab:objective}
  \centering
  \resizebox{0.7\textwidth}{!}{
    \begin{tabular}{ l c c c }
      \toprule
      \textbf{Model} & \textbf{CER(\( \% \))$\downarrow$} & \textbf{SPK-COS$\uparrow$} & \textbf{Latency(ms)$\downarrow$} \\ 
      \midrule
      ground-truth              & 1.54  & N/A    & N/A      \\
      Seed-VC                   & \textbf{2.28}  & 0.611  & N/A    \\
      DualVC2                   & 6.31  & 0.530  & 186.4    \\
      DualVC3(stand-alone mode) & 9.77  & 0.511  & 43.58    \\
      \midrule
      SynthVC-large             & \textbf{6.04}  & \textbf{0.648}   & 96.3    \\
      SynthVC-small             & 8.38  & 0.587   & 57.9    \\
      SynthVC-base              & 6.27  & 0.626   & \textbf{77.1}    \\
      \bottomrule
    \end{tabular}}
\end{table*}

\section{Experiments Results}

\subsection{Subjective and Objective Evaluation}

As shown in Table~\ref{tab:subjective}, SynthVC-base achieves an N-MOS of 3.51, S-MOS of 3.77, and I-MOS of 3.76, outperforming both DualVC2 and DualVC3 
across all subjective metrics. Although Seed-VC achieves the highest scores (e.g., S-MOS of 4.34), it is a diffusion-based \emph{non-streamable} model, and thus not directly applicable in real-time settings.

As shown in Table~\ref{tab:objective}, SynthVC-base achieves a CER of 6.27\% and a SPK-COS of 0.626. 
Interestingly, its speaker similarity (SPK-COS) is slightly higher than that of Seed-VC (0.611), which we attribute to SynthVC being trained to convert speech into a fixed set of target speakers, while Seed-VC performs zero-shot inference across arbitrary speakers.

\subsection{Model Size and Efficiency}

Table~\ref{config} summarizes the model scaling results. SynthVC-small uses only 8.24M parameters and 5.21G MACs per second of audio, making it suitable for edge deployment, though with a trade-off in intelligibility and fidelity. SynthVC-base achieves the best trade-off between performance and efficiency, while SynthVC-large provides the best perceptual quality at higher computational cost. 

\subsection{Streaming Latency Analysis}
 
All latency measurements are conducted on a single-core Intel i5-10210U CPU. For SynthVC, total latency is computed as the sum of chunk size (50\,ms) and measured inference time: 21.9\,ms for SynthVC-small (total 71.9\,ms), 27.1\,ms for SynthVC-base (77.1\,ms), and 46.3\,ms for SynthVC-large (96.3\,ms).

For the baselines, we report latency figures directly from their original papers:

\begin{itemize}
    \item DualVC2: Reports a total latency of 186.4\,ms, composed of a 160\,ms chunk size and 26.4\,ms model inference time.
    \item DualVC3 (stand-alone mode): Combines 3.58\,ms model inference, 20\,ms chunk-waiting, and 20\,ms lookahead buffer, totaling 43.58\,ms.
\end{itemize}

While DualVC3 achieves the lowest latency, it suffers significantly in perceptual metrics. SynthVC-base provides a compelling balance, achieving superior quality with only 77.1\,ms total latency.

\begin{table}[t]
    \centering
    \caption{The configurations of SynthVC with different parameter sizes are as follows. \( H \) denotes the dimension of the latent representation. The Multiply-Accumulate Operations (MACs) indicate the computational result for processing a 1-second segment of input audio.} 
    \begin{tabular}{l|c|c c}
    \hline
        model          & H    & params(M)& MACs(G)   \\ \hline
        SynthVC-base  & 256  & 14.70    & 8.89      \\ \hline
        SynthVC-large & 512  & 57.56    & 35.39     \\ 
        SynthVC-small & 192  & 8.24     & 5.21      \\ \hline
    \end{tabular}
    \label{config}
\end{table}

\section{Ablation Study\label{ablation}}

\begin{figure}[t]
  \centering
  \begin{subfigure}{0.45\textwidth}
    \centering
    \includegraphics[width=\linewidth]{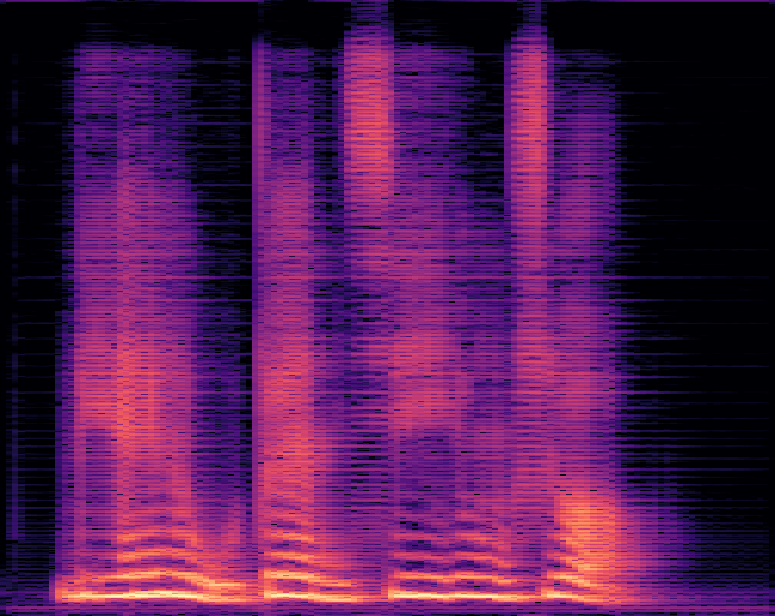}
    \caption{w/o Converter}
    \label{fig:a1}
  \end{subfigure}
  \hfill
  \begin{subfigure}{0.45\textwidth}
    \centering
    \includegraphics[width=\linewidth]{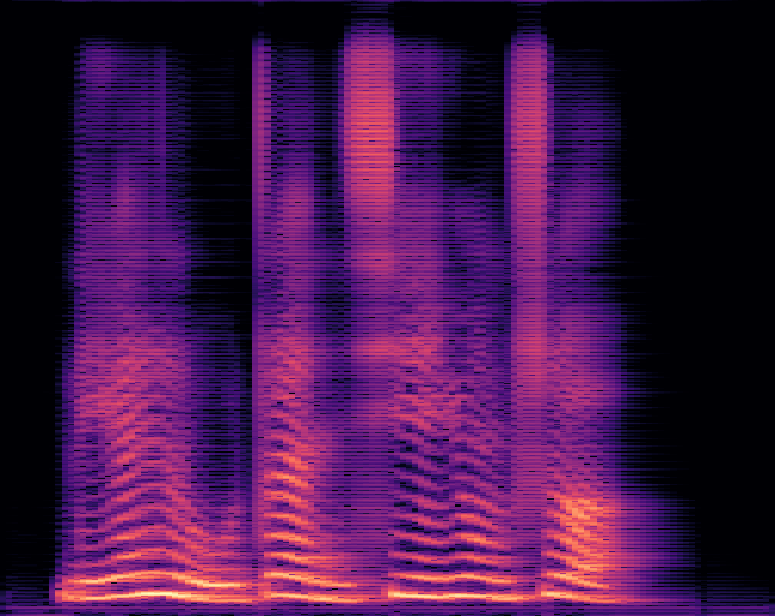}
    \caption{SynthVC}
    \label{fig:dis1}
  \end{subfigure}
  \vskip\baselineskip
  \begin{subfigure}{0.45\textwidth}
    \centering
    \includegraphics[width=\linewidth]{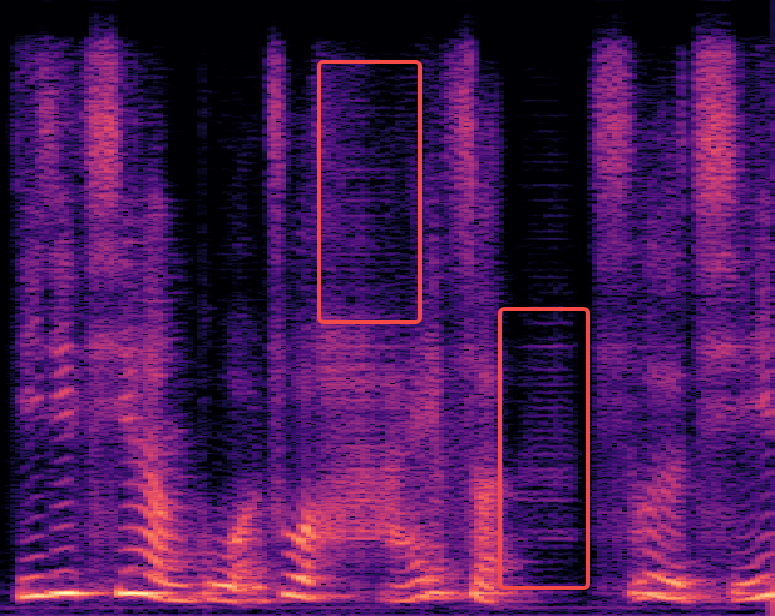}
    \caption{w/o Aligned Training}
    \label{fig:a2}
  \end{subfigure}
  \hfill
  \begin{subfigure}{0.45\textwidth}
    \centering
    \includegraphics[width=\linewidth]{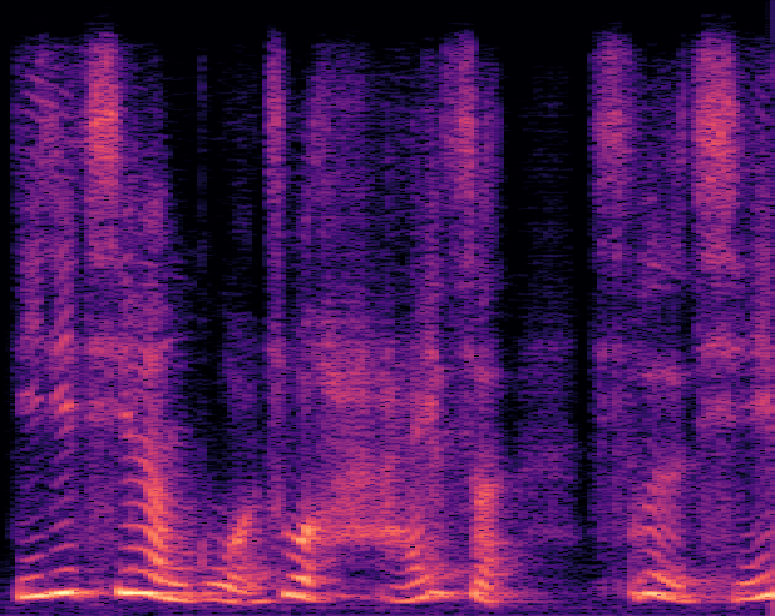}
    \caption{SynthVC}
    \label{fig:dis2}
  \end{subfigure}
  \caption{Spectrogram comparison of converted audio across ablation variants and SynthVC. (a) and (b) show results after only Stage 1 training. Removing the Converter (a) leads to blurred high-frequency details, while SynthVC (b) preserves spectral fidelity. (c) and (d) correspond to Stage 2 training. Omitting alignment between training and inference (c) introduces spectral artifacts, whereas SynthVC (d) eliminates such artifacts and maintains high-frequency detail.} 
  \label{fig:tsne}
\end{figure}

To evaluate the effectiveness of the Converter and Aligned Training, we conduct two ablation experiments, each modifying a key component of SynthVC. Figure~\ref{fig:tsne} illustrates the spectral effects of these ablations, compared to the full SynthVC system.

\subsubsection{Effect of the Converter}
We remove the Converter and directly train the Autoencoder using synthetic waveform pairs. In this setting, the model is expected to learn timbre transformation implicitly from waveform supervision. As shown in Figure~\ref{fig:a1}, this results in over-smoothed outputs with loss of high-frequency spectral detail. This confirms the importance of explicitly modeling speaker transformation in the latent space.

\subsubsection{Effect of Aligned Training}
In this ablation, we directly feed the Decoder with the Encoder's latent output during adversarial training (stage 2), while still using the Converter's output during inference. This setup introduces a mismatch, where the Decoder is exposed to different latent distributions during training and inference. As shown in Figure~\ref{fig:a2}, this inconsistency leads to noticeable spectral artifacts and degraded audio quality. In contrast, SynthVC adopts \textit{Aligned Training}, which uses the Converter's output consistently in both phases, resulting in more natural and stable synthesis.

\section{Conclusions}


We proposed SynthVC, a lightweight end-to-end streaming voice conversion framework with an end-to-end latency of 77.1 ms. By combining a neural codec backbone with synthetic parallel data generated by a zero-shot VC model, SynthVC enables high-quality waveform-to-waveform conversion without the need for ASR-based content features or disentanglement strategies. Extensive experiments demonstrate that SynthVC consistently outperforms baselines streaming VC models in both naturalness and speaker similarity, while remaining efficient enough for real-time deployment.

%
%
%
%

\end{document}